\documentclass[onecolumn,pra,preprint]{revtex4}
\usepackage{epsf}
\usepackage{amsmath,amsthm,amssymb}
\usepackage{color}
\usepackage{dcolumn}
\usepackage{bm}
\usepackage{graphicx,epsfig}
\usepackage[dvipsnames]{xcolor}

\graphicspath{{fig/}}
\everymath{\displaystyle}

\begin{document}           

\title{Electric quadrupole moment and the tensor magnetic polarizability of twisted electrons and a potential for their measurements}

\author{Alexander J. Silenko $^{1,2,3}$} \email{alsilenko@mail.ru}
\author{Pengming Zhang$^{1,4}$}
\email{zhpm@impcas.ac.cn}
\author{Liping Zou$^{1,4}$}
\email{zoulp@impcas.ac.cn}

\affiliation{$^1$Institute of Modern Physics, Chinese Academy of Sciences, Lanzhou 730000, China}
\affiliation{$^2$Bogoliubov Laboratory of Theoretical Physics, Joint Institute for Nuclear Research, Dubna 141980, Russia}
\affiliation{$^3$Research Institute for Nuclear Problems, Belarusian State University, Minsk 220030, Belarus}
\affiliation{$^4$University of Chinese Academy of Sciences, Yuquanlu 19A, Beijing 100049, China}

\date{\today}

\begin{abstract}
For a twisted (vortex) Dirac particle in nonuniform electric and magnetic fields, the relativistic Foldy-Wouthuysen Hamiltonian is derived including high order terms describing new effects.
The result obtained
shows for the first time that a twisted spin-1/2 particle possesses
a tensor magnetic polarizability and a measurable (spectroscopic) electric quadrupole moment. We have calculated the former parameter and have
evaluated the latter one for a twisted electron.
The tensor magnetic polarizability of the twisted electron can be measured in a magnetic storage ring because a beam with an initial orbital tensor polarization acquires a horizontal orbital vector polarization. The electric quadrupole moment is rather large and strongly influences the dynamics of the intrinsic orbital angular momentum. Three different methods of its measurements, freezing the intrinsic orbital angular momentum and two resonance methods, are proposed. The existence of the quadrupole moment of twisted electrons can lead to practical applications.
\end{abstract}

\maketitle


The discovery of twisted (vortex) electron beams \cite{UTV} whose existence was predicted in Ref. \cite{Bliokh2007} has shown that particles can carry an intrinsic orbital angular momentum (OAM). Since twisted electrons possess large magnetic moments, this discovery
opens new possibilities in the electron microscopy and investigations of magnetic phenomena
(see Refs. \cite{BliokhSOI,Lloyd,LloydPhysRevLett2012,Rusz,Edstrom,imaging,Observation,OriginDemonstration}
and references therein). Twisted electron beams with large intrinsic OAMs (up to 1000$\hbar)$
have been recently obtained \cite{VGRILLO}.
Basic properties of free twisted
beams
have been considered in Refs. \cite{Bialynicki-Birula,Barnett}.
The dynamics of the intrinsic OAM in external magnetic and electric fields has been studied in Refs. \cite{Bliokh2007,Bliokhmagnetic,magnetic,Greenshields,classicalmagnetic,experimentmagnetic}. The general relativistic description of the classical and quantum dynamics of the intrinsic OAM in arbitrary electric and magnetic fields has
been discussed in Refs. \cite{Manipulating} and \cite{ResonanceTwistedElectrons}, respectively. In Ref. \cite{ResonanceTwistedElectrons}, the relativistic quantum dynamics of twisted (vortex) electrons has been constructed in the Schr\"{o}dinger form on the basis of the relativistic Foldy-Wouthuysen (FW) transformation \cite{JMP,Classicallimit,relativisticFW}.
In the present work, an application of the approach used in Ref. \cite{ResonanceTwistedElectrons} allows us to obtain new fundamental
properties of twisted electron beams. We demonstrate for the first time that a twisted spin-1/2 particle can possess a large measurable (spectroscopic) electric quadrupole moment (EQM) and a tensor magnetic polarizability (TMP). We calculate these new fundamental parameters
(caused by the intrinsic OAM) for a twisted electron and develop methods for their measurements.

While a twisted electron is a single pointlike particle
described by the standard Dirac equation, its wave function
has a nontrivial spatial structure (see the reviews \cite{BliokhSOI,Lloyd}). A particle with an intrinsic OAM is characterized by
nontrivial solutions of the Dirac and Schr\"{o}dinger equations. Such solutions are coherent superpositions of partial plane waves with different momenta \cite{BliokhSOI,Lloyd,IvanovScattering}.

The relativistic FW transformation (see Refs. \cite{JMP,Classicallimit,relativisticFW} and references therein) being the relativistic generalization of the original method 
\cite{FW} can be applied to obtain the Schr\"{o}dinger form of the relativistic quantum mechanics. The \emph{exact} relativistic Hamiltonian in the FW representation (the FW Hamiltonian) for a twisted or a untwisted Dirac particle $(g=2)$ in a static (in general, nonuniform) magnetic field has been first obtained in Ref. \cite{Case} and is given by $(\hbar=1,~c=1)$ \cite{JMP,Case,Energy1,Energy3}
\begin{equation}
\begin{array}{c}
{\cal H}_{FW}=\beta\sqrt{m^2+\bm{\pi}^2-e\bm\Sigma\cdot\bm B},
\end{array}
\label{eq33new}
\end{equation}
where $\bm{\pi}=\bm{p}-e\bm A$ is the kinetic momentum, $\bm B$
is the magnetic induction, and $\beta$ and $\bm\Sigma$ are the Dirac matrices.
This Hamiltonian acts on the bispinor $\Psi_{FW}=
\left(\begin{array}{c} \phi \\ 0 \end{array}\right)$.

A twisted electron is a charged centroid \cite{Bliokh2007,BliokhSOI}. Needed derivations
are similar to those in Ref. \cite{ResonanceTwistedElectrons}. However, second-order terms in $B$ should be
calculated. We can suppose that the
de Broglie wavelength, $\hbar/p$, is much smaller than the characteristic
size of the nonuniformity region of the external field.
Summing over partial waves with different momentum directions brings the operator $\bm{\pi}^2$ to the form \cite{ResonanceTwistedElectrons}
\[
\begin{array}{c}
\bm{\pi}^2={\bm{\pi}'}^2+{\pmb{\mathfrak{p}}}^2
-\frac e2\left[\bm L\cdot\bm B(\bm R)+\bm B(\bm R)\cdot\bm L\right]
+\frac {e^2}{4}\bigl[\bm B(\bm R)\times\pmb{\mathfrak{r}}\bigr]^2,
\end{array}
\]
where
$\bm R$ and $\bm{\pi}'$ are the center-of-charge radius vector and the kinetic momentum of the centroid as a whole, $\pmb{\mathfrak{r}}=\bm r-\bm R$ and $\pmb{\mathfrak{p}}=-i\hbar\partial/(\partial\pmb{\mathfrak{r}})$ are internal canonical variables, and $\bm L\equiv\pmb{\mathfrak{r}}\times\pmb{\mathfrak{p}}$ is the intrinsic OAM \cite{ResonanceTwistedElectrons}.
The straightforward extraction of the square root \cite{SM1}
brings Eq. (\ref{eq33new}) into the form
\begin{eqnarray}
{\cal H}_{FW}&=&\beta\epsilon
-\beta\frac {e}{4}\left(\frac{1}{\epsilon}\bm\Lambda\cdot\bm B(\bm R)
+\bm B(\bm R)\cdot\bm\Lambda\frac{1}{\epsilon}\right) \nonumber \\
&&+\beta\frac {e^2}{16}\left(\left\{\frac{1}{\epsilon},\bigl[\bm B(\bm R)\times\pmb{\mathfrak{r}}\bigr]^2\right\}-\left\{\frac{1}{\epsilon^3},\bigl[\bm B(\bm R)\bigr]^2\right\}\right) \nonumber \\
&&
-\beta\frac {e^2}{16}\Big ( \frac{1}{\epsilon^3}\bigl[\bm L\cdot\bm B(\bm R)
+\bm B(\bm R)\cdot\bm L\bigr]\bigl(\bm\Sigma\cdot\bm B(\bm R)\bigr)
\nonumber \\
&& +\bigl(\bm\Sigma\cdot\bm B(\bm R)\bigr)\bigl[\bm L\cdot\bm B(\bm R)+\bm B(\bm R)\cdot\bm L\bigr]\frac{1}{\epsilon^3}\Big )
\nonumber \\
&&-\beta\frac {e^2}{64}\left\{\frac{1}{\epsilon^3},\bigl[\bm L\cdot\bm B(\bm R)+\bm B(\bm R)\cdot\bm L\bigr]^2\right\},
\nonumber \\
\epsilon &=& \sqrt{m^2+{\bm{\pi}'}^2+{\pmb{\mathfrak{p}}}^2}.
\quad \bm\Lambda=\bm L+\bm\Sigma.
\label{eqHmsum}
\end{eqnarray}

In Eq. (\ref{eqHmsum}), all terms proportional to $1/\epsilon^3$ were not previously taken into account.
The curly brackets $\{\dots,\dots\}$ denote anticommutators. The additional terms appearing in the second order expansion of the square root in Eq. (\ref{eq33new}) in a power series in $1/\epsilon$ are quadratic or bilinear in $\bm B$.
Usually, $\epsilon^2\sim{m^2}=4.41\times10^9$ $e\cdot$T, $L\sim 100, B\sim1$ T, and these terms are approximately 7 orders less than the main OAM-dependent term. Therefore, our previous results and conclusions \cite{ResonanceTwistedElectrons} are correct, while the additional OAM-dependent terms define new physical effects.
The second to last term in the FW Hamiltonian (\ref{eqHmsum}) characterizes the spin -- intrinsic OAM coupling in the magnetic field and describes the additional spin precession 
caused by the intrinsic OAM. Reversing $\bm L$ changes the sign of the OAM-dependent correction to the spin precession frequency.
The existence of the spin -- intrinsic OAM coupling has been previously established in Refs. \cite{BliokhSOI,Lloyd,Bliokhmagnetic,magnetic}.
The last term describing the tensor interaction of the intrinsic OAM with the magnetic field
is similar to the corresponding spin tensor interaction. The
operator of the latter interaction has the form $W=-\beta_T(\bm S\cdot\bm B)^2$, where $\beta_T$ is the TMP defined in the particle rest frame and $\bm S$ is the spin matrix $[\bm S^2=s(s+1)]$.
Thus, the TMP
caused by the intrinsic OAM is given by
\begin{equation}
W=-\beta\beta_T(\bm L\cdot\bm B)^2,\quad \beta_T=\frac {e^2\hbar^2}{8m^3}=5.25\times10^4\,{\rm fm}^3.
\label{eqTMP}
\end{equation}
The TMP of the twisted electron is much larger than TMPs of particles and nuclei conditioned by the spin interactions. In particular, for pointlike
W$^\pm$ bosons $\beta^{(W)}_T\sim10^{-11}$ fm$^3$
\cite{PhysRevDspinunit,PhysRevDunitexact}.
For the deuteron, the theoretical estimation is $\beta_T=0.195$ fm$^3$ \cite{BETA_T}. In addition, the tensor interaction of the twisted electron is proportional to $L^2\sim10^4 - 10^6$.

The operator $\bm L$ commutes with $\pmb{\mathfrak{p}}^2$. The noncommutation of this operator with $\bigl[\bm B(\bm R)\times\pmb{\mathfrak{r}}\bigr]^2$ does not lead to any important effects because the expectation values of the nonzero commutators $\left[L_i,(\bm B(\bm R)\times\pmb{\mathfrak{r}})^2\right]\,(i=x,y$ when $\bm B(\bm R)=B\bm e_z)$ are \textbf{equal} to zero.

Thus, the evolution of the intrinsic OAM does not reduce to its precession. The same assertion has been made in Refs. \cite{BliokhSOI,Lloyd,magnetic,classicalmagnetic}. However, Eq. (\ref{eqHmsum}) shows the existence of a new
interaction caused by the TMP of the twisted Dirac particle.
The corresponding relativistic classical equation has the same form as Eq. (\ref{eq33new}) (except for the spin term).
Moreover, a consideration of a twisted centroid leads to the \emph{classical} equation similar to Eq. (\ref{eqHmsum}). The tensor electric and magnetic polarizabilities caused by the spin interactions are 
common properties of nuclei with the spin $s\ge1$. Besides this, the exact FW Hamiltonians for pointlike spin-1/2 and spin-1 particles (with $g=2)$ in a uniform magnetic field (see Eq. (\ref{eq33new}) and Refs. \cite{PhysRevDspinunit,PhysRevDunitexact}) are
almost identical (the only difference is the form of the spin matrices). The existence of the TMP
for the pointlike spin-1 particles additionally substantiates its existence for the twisted Dirac ones. We also
mention
that the three-component spin operator $\bm s=\hbar\bm\Sigma/2$ and the OAM operator $\bm L$ are defined in the particle rest frame and in the lab frame, respectively.

While the TMP of the twisted electron is large as compared with that of the deuteron, its measurement is a difficult experimental task. The TMP leads to very small shifts of 
energy levels and
cannot be determined using the magnetic-resonance method. While $(B_\|\cos{\omega t})^2=B_\|^2(1+\cos{2\omega t})/2$, an oscillating horizontal $(\|)$ magnetic field $\bm B_\|$ which frequency is half that of transitions between the Landau levels cannot stimulate any resonance. This occurs because the operator $(\bm B_\|\cdot\bm L)^2$ (unlike the operator $\bm B_\|\cdot\bm L)$ does not mix neighboring energy levels. The same situation takes place for the tensor polarizabilities caused by the spin. In particular, the operators $(\bm E_\|\cdot\bm S)^2$, $(\bm B_\|\cdot\bm S)^2$ mix only the levels $s_z=\pm1$ for spin-1 particles \cite{PRC2007,PRC2008,PRC2009,JPhysConfSer}.

The best possibilities to measure the TMP of the twisted electron are provided by the effects found in Ref. \cite{Bar3} and investigated in detail in Refs. \cite{PRC2008,BarJPhysG}. If the TMP is caused by the spin interactions, it produces a spin rotation with two frequencies instead of one, beating with a frequency proportional to $\beta_T$, and 
transitions between vector and tensor polarizations \cite{Bar3,BarJPhysG}. Following Ref. \cite{PRC2008}, we propose to use a tensor-polarized twisted electron beam in a magnetic storage ring. In this case, the TMP is the only reason of the appearance of a horizontal orbital vector polarization of the beam. This polarization grows 
almost linearly in time \cite{SM1,PRC2008}. The horizontal component of the intrinsic OAM rotates with the Larmor frequency.
The experiment can be performed in an electron storage ring or in a Penning trap. It needs a high beam coherency. To reach such a coherency, some methods developed for the electric-dipole-moment experiment \cite{PhysRevLett2016} can be applied.

We also consider OAM-dependent interactions proportional to field derivatives.
Contrary to the pointlike electron, the twisted electron (centroid) has a highly anisotropic spatial structure. Such an object
possesses the EQM, while this property has not been previously mentioned. The Laguerre-Gaussian wave function describing a wave beam \cite{BliokhSOI,Lloyd} does not allow a rigorous determination of the electron density shape \emph{in the centroid rest frame} $(p_z\rightarrow0)$. It is natural to assume that this is a strongly oblate (pancake
shaped) spheroid. Twisted electron states in a uniform magnetic field \cite{BliokhSOI,Bliokhmagnetic} always have such a shape. In this case, the intrinsic EQM of the centroid is given by $(e=-|e|)$
\begin{equation}
Q_0=-e<{\mathfrak{r}}^2>=-e\frac{\int{\varrho(\mathfrak{r}){\mathfrak{r}}^3d\mathfrak{r}}}{\int{\varrho(\mathfrak{r})\mathfrak{r} d\mathfrak{r}}},
\label{eqQM}
\end{equation} where $\varrho(\mathfrak{r})=|\Psi_{FW}|^2$ is the electron density and $\mathfrak{r}$ is the radial coordinate of the cylindrical coordinate system introduced relative to the center of charge of the centroid. In the nonrelativistic approximation, the square of the FW wave function 
reduces to the square of the corresponding Schr\"{o}dinger wave function.

The interaction of an extended charged particle with a static electric field is defined by the operator
\begin{equation}
W=-\frac12\bigl[\bm d\cdot\bm E(\bm R)+\bm E(\bm R)\cdot\bm d\bigr]-\frac{1}{12}\left\{Q_{ij},\frac{\partial E_i(\bm R)}{\partial X_j}\right\}, \label{eqW}
\end{equation} where $\bm d$ and $Q_{ij}$ are the operators of the electric dipole moment (see Ref. \cite{ResonanceTwistedElectrons}) and the EQM, respectively. In the centroid rest frame, the EQM interacts only with an
electric field.
The EQM operator of the twisted electron averaged on states with the specific total angular momentum $\bm j=\bm L+\bm s$ is defined in the \emph{centroid rest frame} and is given by
\begin{equation}
Q_{ij}=\frac{3Q_s}{2j(2j-1)}\left[\{j_i,j_j\}-\frac23\delta_{ij}j(j+1)\right],
\label{QuMomentOper} \end{equation} where $Q_s$ is the \emph{spectroscopic} EQM. Its connection with the intrinsic EQM has the form \cite{EQM,EQMreview}
\begin{equation}
Q_s=\frac{3K^2-j(j+1)}{(j+1)(2j+3)}Q_0,
\label{EQMim} \end{equation}
where $K$ is the projection of the total angular momentum onto the symmetry axis of the particle. In the case at hand, $j\gg1$ and $Q_s\approx Q_0$.

For the Landau problem, eigenfunctions of the nonrelativistic and relativistic Hamiltonians coincide \cite{ResonanceTwistedElectrons}. For twisted and untwisted electrons, the mean square of the radial cylindrical coordinate of \emph{the pointlike electron} is given by \cite{Bliokhmagnetic}
\begin{equation}
<r^2>=\frac{w_m^2}{2}\left(2n+|l_z|+1\right),\qquad w_m=2\sqrt{\frac{\hbar}{|eB|}}.
\label{eqrquar} \end{equation}
Here
$w_m$ is the beam waist \cite{BliokhSOI,Bliokhmagnetic} and $\bm l$ is the sum of the intrinsic and extrinsic OAMs. When $B=1$ T, $w_m=5.1\times10^{-8}$ m.

The diameter of the vortex beam depending on the OAM has been determined in Refs. \cite{imaging,OAMdiameter}. It
is about 10 nm when the topological charge is $\mathfrak{m}=50$ $(L_z=\mathfrak{m}\hbar)$ and is proportional to $\mathfrak{m}$ \cite{OAMdiameter}.

The spin also contributes to the EQM of the twisted electron. An orbital motion of the magnetic moment of a spinning particle leads to the appearance of an electric current quadrupole moment (ECQM) \cite{ECQM}:
\begin{equation}
Q_{ij}^{(curr)}=-\frac{1}{2\epsilon}\left[3L_i\mu_j+3L_j\mu_i-2\delta_{ij}(\bm L\cdot\bm\mu)\right],\quad \bm\mu=\frac{e\bm s}{\epsilon}.
\label{ECQM} \end{equation} Its appearance results in a small correction to the spin precession frequency. The sign of this correction is defined by the sign of $L_z$. The ECQMs are comparatively small $(Q^{(curr)}\sim eL/m^2)$. The correction to the spin
precession frequency is of the following order:
$$\Delta\Omega\sim L\left|\frac{\partial E_i}{\partial X_j}\right|_{max}\times10^{-10}\,{\rm s}^{-1},$$
where the maximum gradient of the electric field is given in units of V/m$^2$. The
ECQMs cause a spin -- intrinsic OAM coupling.

Let us analyze the potential for measuring
the EQM of twisted electrons in storage rings when $L\gg1$ and spin effects are neglected.
It is convenient to determine the dynamics of $L_z$. Relativistic effects in interactions of EQMs of spinning particles with electric and magnetic fields have been described in Refs. \cite{PKS,MovingQ,Spin1JETP}. 
Since polarization effects conditioned by the spin and intrinsic OAM are similar \cite{SM1}, we can use the results \cite{PKS,MovingQ,Spin1JETP} with substituting $\bm L$ for $\bm S$ and
taking into account that $L>>1$
and $\bm V\cdot\bm E=0$ $(\bm V$ is the centroid velocity operator).
The
quadrupole interaction in the lab frame
reads $(\bm{\widetilde{\beta}}=\bm V/c=\widetilde{\beta}\bm e_\phi)$
\begin{eqnarray}
W&=&-\frac{Q_s}{4L^2}\left[(\bm L\cdot\nabla)-
\frac{\gamma}{\gamma+1}(\bm L\cdot\bm{\widetilde{\beta}})(\bm{\widetilde{\beta}}\cdot\nabla)\right]
(\bm L\cdot\bm{\mathcal{E}}), \nonumber \\
\bm{\mathcal{E}}&=&\bm E(\bm R)+\frac12\left[\bm{\widetilde{\beta}}\times\bm B(\bm R)-\bm B(\bm R)\times\bm{\widetilde{\beta}}\right],
\label{Equadmn}
\end{eqnarray}
where $\bm{\mathcal{E}}=\bm{\mathcal{E}}(\bm R)$ is the quasielectric field. All fields are defined in the lab frame. 
The noncommutativity of operators is neglected. The intrinsic OAM presented in Eq. (\ref{Equadmn}) should not be confused with the extrinsic OAM.
The operator $W$ should be added to the Hamiltonian (\ref{eqHmsum}).
The use of a nonuniform magnetic field
for focusing may be preferable.


The interaction operator $W$ contains the terms proportional to
$\partial\mathcal{E}_z/\partial Z$ and $\partial\mathcal{E}_r/\partial R$ $(R$ is the radial coordinate of the cylindrical coordinate system). The former term commuting with $L_z$ can be disregarded. The effect of the latter term is defined by
\begin{equation}
W=-\frac{Q_s}{4L^2}L_r^2\frac{\partial\mathcal{E}_r}{\partial R}.
\label{Equadmm}
\end{equation}
The last term in Eq. (\ref{QuMomentOper}) does not contribute to the interaction operator (\ref{Equadmm}) because $\partial\mathcal{E}_i/\partial X_i\equiv{\rm div}\,\bm{\mathcal{E}}=0$.

To determine the effect of $W$ on the dynamics of the intrinsic OAM, we can use the fact that the components of the spin and the OAM satisfy equivalent commutation relations. Therefore,
the OAM polarization tensor $\{L_i,L_j\}$ rotates in external fields with the same angular velocity as the OAM (the spin polarization tensor possesses the equivalent property \cite{JPhysG2015}).
The dynamics of the intrinsic OAM is defined by the large term \cite{ResonanceTwistedElectrons}
\begin{eqnarray}
\frac{d\bm L}{dt}&=&\frac12
\left(\bm\Omega\times\bm L-\bm L\times\bm\Omega\right), \nonumber \\
\bm\Omega &=& -\beta\frac{e}{4}\left\{\frac{1}{\epsilon},\bm B(\bm R)\right\}
+
\frac{e}{4}\left[\frac{1}{\epsilon^2}\bm\pi'\times\bm E(\bm R)-\bm E(\bm R)
\times\bm\pi'\frac{1}{\epsilon^2}\right]
\label{DynaOAM}
\end{eqnarray}
and by corrections to this term caused by the EQM.
Other corrections are defined by Eqs. (\ref{eqHmsum}) and (\ref{eqTMP}).
In Eq. (\ref{DynaOAM}), $\bm\Omega$ is the operator of the angular velocity of Larmor precession in external fields. The Larmor precession caused by the vertical magnetic field and by the radial electric one (if the latter field is also used) does not change $L_z$.

In Eqs. (\ref{eqHmsum}), (\ref{Equadmm}), and (\ref{DynaOAM}), a noncommutativity of the operators of coordinate and intrinsic OAM can be ignored because the operators $\pmb{\mathfrak{p}}$ and $\bm L$ are defined only by the internal coordinates.

The change of $L_z$ caused by the quadrupole interaction (\ref{Equadmm}) is observable
only when the Larmor precession is eliminated. This can be done by \emph{freezing the intrinsic OAM} \cite{Manipulating} in
a specific combination of vertical magnetic and radial electric fields equalizing the angular velocities of the beam rotation and the intrinsic-OAM one $(\bm\omega=\bm\Omega)$. An angle between the intrinsic OAM and the momentum remains unchanged. A similar method of freezing the \emph{spin} \cite{frozenEDM} may be applied in electric-dipole-moment 
experiments.
Freezing the intrinsic OAM takes place when \cite{Manipulating}
\begin{equation} \bm B_0=\left(\frac{2}{{\widetilde{\beta}}^2}-1\right)\bm{\widetilde{\beta}}\times\bm E,\qquad \bm\omega=-\frac{e\bm B_0}{mc\gamma(\gamma^2+1)}.
\label{eqBA}\end{equation} If magnetic focusing is used, the magnetic field $\bm B$ is nonuniform and $\bm B_0$ is an average magnetic field. The forces caused by the electric and magnetic fields are oppositely directed. Electrons move counterclockwise. The
ring radius
is defined by
\begin{equation} R_0=\frac{V}{\omega}=(\gamma^2+1)\sqrt{\gamma^2-1}\,\frac{mc^2}{|e|B_0}.
\label{ringr}\end{equation}
The nonuniformity of  $\bm B$
leads to a nonuniform electric field in the centroid
rest frame and to a turn of the intrinsic OAM in the vertical plane. In this case,
\begin{equation} \frac{\partial B_z}{\partial R}=-\frac{nB_0}{R_0},\qquad 0<n<1,
\label{eqBn}\end{equation} where $n$ is the field index. In Eq. (\ref{Equadmm}), $\mathcal{E}_r=\widetilde{\beta}B_z$ and $\partial\mathcal{E}_r/\partial R=-\widetilde{\beta}nB_0/R_0$.

The commutator of the total FW Hamiltonian (including
$W)$ with the OAM operator results in the following addition to the equation of motion:
\begin{equation}
\begin{array}{c}
\frac{dL_z}{dt}=\frac{Q_s\widetilde{\beta}nB_0}{4L^2R_0}\{L_r,L_\phi\}.
\end{array}
\label{eqmotion}
\end{equation}
Therefore, a beam with an initial horizontal orbital polarization acquires a vertical orbital polarization (cf. the similar spin effect \cite{PRC2007,PRC2008,PRC2009,JPhysConfSer,Bar3,BarJPhysG,Bar1}).
The beam can be tensor-polarized (when two beams with opposite orbital polarizations are joined) or vector-polarized.
Let $\psi$ be the angle defining the orbital polarization relative to the $\bm e_r$ and $\bm e_\phi$ axes. The azimuth $\psi=0$ characterizes the intrinsic OAM directed radially outward. The change of $L_z$ is maximum when the direction of the initial horizontal orbital polarization satisfies the condition $|\sin{2\psi}|=1$ (cf. Refs. \cite{SM1,PRC2007,PRC2009}).

The effect of the EQM on the OAM dynamics
is very strong and can be easily observed.
When the beam energy is equal to 300 keV and $R_0=0.5$ m, $\widetilde{\beta}=0.777$, $E=2.46$ MV/m, $B_0=0.0148$ T, $f=\omega/(2\pi)=7.41\times10^7$ Hz.
When $L\sim100$ and the OAM diameter is determined based on the data presented in Ref. \cite{OAMdiameter}, the quantity $Q_s/(|e|R_0)\sim10^{-16}$ m is not negligible as compared with the reduced Compton wavelength of the electron ${\lambdabar}_C=\hbar/(mc)=3.86\times10^{-13}$ m. 
The frequency of the cyclic evolution of the orbital polarization (cf. Refs. \cite{PRC2007,PRC2008,PRC2009,Bar3,BarJPhysG,Bar1}) is 5 orders of magnitude less than the cyclotron frequency $f$ and can be properly measured. The corresponding terms in the FW Hamiltonian also
differ by 5 orders of magnitude. Therefore, the main intrinsic-OAM dynamics is correctly described by the equations obtained in Refs. \cite{Manipulating,ResonanceTwistedElectrons}, while the new EQM-dependent effect is rather important.

There is a systematical error caused by the small \emph{vertical} electric field and the corresponding radial magnetic field $<B_r>=-<E_z>/\widetilde{\beta}$ leading to a vanishing average Lorentz force (cf. Ref. \cite{frozenEDM}).
This systematical error originates from field misalignments. However, it seems to be small and can be eliminated in measurements at two values of the field index.

The EQM of the twisted electron can also be measured by the \emph{magnetic-resonance method}. In this case, a constant electric field is not needed. The resonance effect is provided by a \emph{nonuniform} field oscillating with the angular frequency $\Omega$. The resonance field vanishing in the center of the beam trajectory, $B_z=\mathfrak{B}(R-R_0)\cos{(\Omega t+\varphi)}$ or $E_r=\mathfrak{E}(R-R_0)\cos{(\Omega t+\varphi)}$, is preferable. It is well known that such a field is equivalent to two fields rotating with the angular velocities $\bm\Omega$ [see Eq. (\ref{DynaOAM})] and $-\bm\Omega$. The first of these creates the resonance effect. In the frame rotating with the angular velocity $\bm\Omega$, the intrinsic-OAM dynamics is similar to that when the intrinsic OAM is frozen \cite{SM1}.

The magnetic-resonance method provides the less sensitivity than the method of freezing the intrinsic OAM because the resonance field usually covers a small part of the ring circumference. Nevertheless, the effect of the EQM on the intrinsic-OAM dynamics can be properly detected. Otherwise, the magnetic-resonance method allows one to apply a stronger magnetic field, a smaller ring size, and, therefore, a lower number of twisted electrons. Another advantage of this method is a simpler experimental setup.

The third method of measuring the EQM of the twisted electron is based on a standard stimulation of resonance transitions by an oscillating longitudinal magnetic field. This method, unlike the
previous one, is sensitive to the quadrupole splitting of Landau levels
defined by Eq. (\ref{Equadmm}). The splitting is caused by the focusing magnetic field creating a nonuniform electric field in the electron
rest frame. Therefore, the quadrupole splitting is proportional to the field index $n$. The stimulating magnetic field can be conditioned by a usual rf cavity and is longitudinal because such a field does not affect the beam motion.
The considered method is similar not only to the magnetic-resonance method but also to the nuclear-quadrupole-resonance one.
The resonance frequencies
defined by a quadrupole structure of the energy levels
depend on $n$.

The three methods considered need an increase in the currently available beam intensity. However, similar experiments can be carried out with a single twisted electron in a Penning trap.

Large EQMs of twisted electrons rather strongly interact with nonuniform electric fields.
We expect that the twisted electrons can be successfully used not only in investigations of magnetic properties (see Refs. \cite{BliokhSOI,Lloyd,LloydPhysRevLett2012,Rusz,Edstrom,Observation,OriginDemonstration}
and references therein) but also for nanoscale measurements of nonuniform electric fields in matter.

In this Letter, we have calculated minor terms in the relativistic FW Hamiltonian describing a twisted Dirac particle in nonuniform electric and magnetic fields.
The results
presented by Eqs. (\ref{eqHmsum}) -- (\ref{QuMomentOper}) have shown for the first time that the twisted electron $(s=1/2)$
possesses a TMP and a spectroscopic EQM. We have calculated the former parameter and have
evaluated the latter one.
It is still generally accepted that only particles and nuclei with spin $s\ge1$
are characterized by these parameters.
The TMP of the twisted electron is several orders of magnitude bigger than that of the deuteron. It can be measured in a magnetic storage ring because a beam with the initial orbital \emph{tensor}
polarization acquires a horizontal orbital \emph{vector} polarization. The EQM is rather large and strongly influences the dynamics of the intrinsic OAM. We propose three different methods of its measurements, freezing
the intrinsic OAM and two resonance methods. We expect that the existence of the EQM of twisted electrons can find practical applications because the EQM interaction with nonuniform electric fields in matter depends on the intrinsic-OAM direction. All the considered effects also take place for twisted positrons. Additional explanations are presented
in the Supplemental Material \cite{SM1}.
\smallbreak

This work was supported by the Belarusian Republican Foundation for Fundamental Research
(Grant No. $\Phi$18D-002), by the National Natural Science Foundation of China (Grants No. 11575254 and No. 11805242),
by the National Key Research and Development Program of China (No. 2016YFE0130800),
and
by the Heisenberg-Landau program of the German Federal Ministry of Education and Research (Bundesministerium f\"{u}r Bildung und
Forschung).
A. J. S. also acknowledges hospitality and support by the Institute of Modern
Physics of the Chinese Academy of Sciences. The authors are grateful to
I. P. Ivanov and O. V. Teryaev for helpful exchanges.


\pagebreak

\onecolumngrid
\begin{center}
  \textbf{\large Supplemental Material to ``Electric quadrupole moment and the tensor magnetic polarizability of twisted electrons and a potential for their measurements''   }\\[.2cm]
\end{center}

\setcounter{equation}{0}
\setcounter{figure}{0}
\setcounter{table}{0}
\setcounter{page}{1}
\renewcommand{\theequation}{S\arabic{equation}}
\renewcommand{\thefigure}{S\arabic{figure}}
\renewcommand{\bibnumfmt}[1]{[S#1]}
\renewcommand{\citenumfont}[1]{#1}

In this Supplemental Material, we specify a derivation of Eq. (2) in our Letter and present
some additional explanations for internal-OAM dynamics of twisted electrons caused by interactions bilinear in the internal OAM $\bm L$.

When some operators $\mathcal A$ and $\mathcal B$ do not commute, their double commutators can be neglected, and the operator $\mathcal B$ is comparatively small,
\begin{eqnarray}
\sqrt{\mathcal A+\mathcal B}=\frac12\left\{\sqrt{\mathcal A(1+x)}\right\}=\frac12\left\{\sqrt\mathcal A,\sqrt{1+x}\right\}=\frac12\left\{\sqrt\mathcal A,\left(1+\frac x2-\frac{x^2}{8}\right)\right\},
\nonumber
\end{eqnarray}
where $x=\frac12\{\mathcal A^{-1},\mathcal B\}$. With the equivalent assumptions, the extraction of the square root allows us to pass from Eq. (1) to Eq. (2) in the Letter.

The explanations for internal-OAM dynamics of twisted electrons
are based on previous studies of spin dynamics of deuterons caused by interactions bilinear in the spin \cite{Bar1,Bar3,BarJPhysG,PRC2007,PRC2008,PRC2009,JPhysConfSer,JPhysG2015}. Since these studies have been performed for deuterons, some of them \cite{PRC2007,PRC2008,PRC2009} have used the fixed spin number $s=1$. Nevertheless, it is well-known that the use of the Pauli spin matrices describing spin-1/2 particles allow one to determine the spin rotation of particles with any spins. Similarly, the application of results obtained in Refs. \cite{PRC2007,PRC2008,PRC2009} gives a right qualitative description of the spin dynamics of particles and nuclei with spins $s>1$ and the intrinsic-OAM dynamics. Therefore, the results obtained in Refs. \cite{Bar1,Bar3,BarJPhysG,PRC2007,PRC2008,PRC2009,JPhysConfSer} with different quantum-mechanical and classical approaches are in the best compliance. Moreover, a correction needed for a description of the intrinsic-OAM dynamics usually reduces to the change $\bm S\rightarrow\bm L$ and an addition of the factor $2L-1$.

There is a deep similarity between polarization effects conditioned by the intrinsic OAM $\bm L$ and the spin $\bm S$ caused by the equivalent commutation relations between the components of these operators:
$$[L_i,L_j]=ie_{ijk}L_k,\qquad [S_i,S_j]=ie_{ijk}S_k,$$ where the square brackets denote commutators. The polarization vectors and the polarization tensors for the OAM and the spin are also equivalent:
\begin{eqnarray}
P
_i&=&\frac{<L_i>}{L},\qquad P
_{ij}=\frac{3<L_iL_j+L_jL_i>-2L(L+1)\delta_{ij}}{2L(2L-1)},\nonumber \\
P
_i&=&\frac{<S_i>}{s},\qquad P_{ij}=\frac{3<S_iS_j+S_jS_i>-2s(s+1)\delta_{ij}}{2s(2s-1)},
\label{PMmu}
\end{eqnarray}
where $P_{ij} = P_{ji}$ and $P_{\rho\rho}+ P_{\phi\phi}+ P_{zz} = 1$. In the considered
case, $i, j$ denote projections onto the axes of the cylindrical
coordinate system. This similarity allows one to apply formulas defining the spin dynamics for a description of the intrinsic-OAM dynamics. Equation (\ref{PMmu}) shows that results obtained in spin physics are useful for the description of dynamics of the intrinsic OAM.

The interaction Hamiltonian depending on the intrinsic OAM can be presented in the form
\begin{equation}
\begin{array}{c}
{\cal H}=\bm\Omega\cdot\bm L+\frac12\alpha_{ij}(L_iL_j+L_jL_i),
\end{array}
\label{SuppMHm}
\end{equation}
where
$\bm\Omega$ is the angular velocity of the Larmor precession of the intrinsic OAM in the cylindrical coordinate system. The equation of the intrinsic-OAM motion has the form
\begin{equation}
\begin{array}{c}
\frac{dL_k}{dt}=(\bm\Omega\times\bm L)_k+\frac12\alpha_{ij}\bigl(e_{kil}\{L_l,L_j\}+e_{kjl}\{L_l,L_j\}\bigr).
\end{array}
\label{SuppMmu}
\end{equation}

This equation shows that the minor second term in Eq. (\ref{SuppMmu}) defines the change of the orbital \emph{vector} polarization depending on the orbital \emph{tensor} polarization. This effect allows one to obtain an orbital vector polarization of an initially tensor-polarized and \emph{vector-unpolarized} twisted beam (cf. the similar spin effects \cite{PRC2007,PRC2008,PRC2009}). Such an initial polarization can be reached by mixing two twisted beams with antiparallel intrinsic OAMs. The second term in Eq. (\ref{SuppMmu}) also influences the dynamics of the orbital vector polarization of an initially vector-polarized twisted beam.

An additional equation for $dP_{ij}/(dt)$
completes the system of equations defining the intrinsic-OAM dynamics. Such a system of equation has been constructed in Refs. \cite{Bar1,Bar3,BarJPhysG} for a description of spin effects.
When
the effect of the second term in the interaction Hamiltonian (\ref{SuppMHm}) on the dynamics of the tensor polarization can be neglected,
the tensor polarization is constant in the frame rotating with the angular velocity $\bm\Omega$. As a result, the first term in the interaction Hamiltonian conditions a rotation of the tensor polarization with this angular velocity \cite{JPhysG2015}. The neglect of the influence of the second term on the dynamics of the tensor polarization allows one to consider only Eq. (\ref{SuppMmu}). However, a change of this dynamics caused by the second term in Eq. (\ref{SuppMHm}) can also be important.

Components of the polarization vector and the polarization tensor of vector- and tensor-polarized twisted beams are defined as well as in the spin physics. When the initial twisted beam is vector-polarized and the
direction of its orbital polarization is defined by the spherical angles
$\theta$ and $\psi$,
\begin{equation}
\begin{array}{c}
P_{\rho}(0)=\sin{\theta}\cos{\psi}, ~~~
P_{\phi}(0)=\sin{\theta}\sin{\psi}, ~~~ P_{z}(0)=\cos{\theta},
\end{array}
\label{eq3}
\end{equation}
\begin{equation}
\begin{array}{c}
P_{\rho\rho}(0)=\frac12\left(3\sin^2{\theta}\cos^2{\psi}-1\right),~~~
P_{\phi\phi}(0)=\frac12\left(3\sin^2{\theta}\sin^2{\psi}-1\right),\\
P_{zz}(0)=\frac12\left(3\cos^2{\theta}\cos^2{\psi}-1\right), ~~~
P_{\rho\phi}(0)=\frac34\sin^2{\theta}\sin{(2\psi)}, \\ P_{\rho z}(0)=\frac34\sin{(2\theta)}\cos{\psi}, ~~~
P_{\phi z}(0)=\frac34\sin{(2\theta)}\sin{\psi}.
\end{array}
\label{eq4}
\end{equation}
When the direction of an initial tensor polarization of a twisted beam obtained by mixing two beams with antiparallel intrinsic OAMs is defined by the spherical angles $\theta$ and $\psi$,
\begin{equation}
\bm P(0)=0
\label{eqPz}
\end{equation}
and the components of the polarization tensor are the same.

Equation (3) in our Letter is equivalent to the corresponding equation describing the spin-dependent tensor magnetic polarizability. The presence of the matrix $\beta$ in Eq. (3) is not important because the lower Foldy-Wouthuysen spinor is zero. As a result, the effects of the polarization vector rotation with two frequencies instead of one, beating with a
frequency proportional to $\beta_T$, and transitions between vector and tensor polarizations \cite{Bar3,BarJPhysG,PRC2008,PRC2009} also take place for twisted beams with an orbital polarization. All formulas obtained in these works remain applicable \cite{footnote}. As follows from the results obtained in Refs. \cite{PRC2008,PRC2009}, the above-mentioned vector-unpolarized twisted beam with the orbital tensor polarization defined by Eq. (\ref{eq4}) acquires the orbital vector polarization. 
When the uniform magnetic field is parallel to the $z$ axis,
\begin{eqnarray}
{\cal H}&=&\bm\Omega\cdot\bm L-\beta_{T}B^2L_z^2,\qquad
\nonumber \\
\dfrac{dL_\rho}{dt}&=&(\bm\Omega\times\bm L)_\rho+\beta_{T}B^2\{L_\phi,L_z\},
\nonumber 
\\ 
\dfrac{dL_\phi}{dt}&=&(\bm\Omega\times\bm L)_\phi- \beta_{T}B^2\{L_\rho,L_z\}.
\label{Sup}
\end{eqnarray}

Equations (\ref{PMmu}) and (\ref{eq4}) show that the orbital tensor polarization should not be horizontal. Its optimal vertical direction is defined by the angle $\theta=\pi/4$. The intrinsic-OAM dynamics is given by formulas obtained in Refs. \cite{Bar3,BarJPhysG,PRC2008,PRC2009}. For a twisted particle with $L=1$ and the above-mentioned orbital tensor polarization, the general equation defining this dynamics has the form (see Refs. \cite{PRC2008,PRC2009})
\begin{eqnarray}
P_\rho(t)&=&-\frac12\sin{(2\theta)}\sin{(\Omega t+\psi)}\sin{(bt)},\qquad \nonumber \\
P_\phi(t)&=&\frac12\sin{(2\theta)}\cos{(\Omega t+\psi)}\sin{(bt)}, \nonumber \\
P_{z}(t)&=&0, \qquad b=-\beta_{T}B^2.
\label{etpwi}
\end{eqnarray}
The final vector polarization is horizontal and its absolute value increases.
When $L>1$, Eq. (\ref{etpwi}) is valid up to a constant factor. The electric quadrupole moment (EQM) of a particle affects the intrinsic-OAM dynamics only when the magnetic field is nonuniform.

The effect of the EQM of twisted electrons on the dynamics of the intrinsic OAM also needs some explanations. When the method of freezing the intrinsic OAM is used, the angular velocity of the Larmor precession is vanished and the intrinsic-OAM motion is defined by Eq. (16) in our Letter. A motion of the intrinsic OAM in the case of $\bm\Omega=0$ is very similar to a spin behavior when the frozen spin method is used. The general description of this behavior is presented in Ref. \cite{PRC2009}. We can use Eqs. (14), (15), and (19) from this article. In our case, with the same denotations as above, the dynamics of the vertical component of the orbital polarization is defined by
\begin{equation}
\begin{array}{c}
P_{z}(t)=\cos{(2{\cal A}t)}\cos{\theta}+\frac12\sin^2{\theta}\sin{(2{\cal A}t)}\sin{(2\psi)},
\end{array}
\label{propv}
\end{equation}
when the initial beam of twisted electrons is vector-polarized and
\begin{equation}
\begin{array}{c}
P_{z}(t)=\frac12\sin^2{\theta}\sin{(2{\cal A}t)}\sin{(2\psi)},
\end{array}
\label{prop}
\end{equation}
when this beam is tensor-polarized. Here \begin{equation}
{\cal A}=-\frac{Q_s\widetilde{\beta}nB_0}{8L^2R_0}.\label{defqA}
\end{equation}
 The tensor magnetic polarizability does not influence $P_z$ \cite{PRC2009}.

When the magnetic-resonance method of a measurement of the EQM is used, the intrinsic OAM rotates with the angular velocity of the Larmor precession, $\bm\Omega$. The angular frequency of the spin precession, $\Omega'$, is defined by the well-known Thomas-Bargmann-Michel-Telegdi equation. Precedent studies of the spin-tensor effects fulfilled by different methods \cite{Bar3,BarJPhysG,PRC2007} have shown that the spin behavior perturbed by time-independent spin-vector and spin-tensor interactions is different. In the same static fields, the former and latter interactions lead to spin oscillations with the angular frequencies $\Omega'$ and $2\Omega'$, respectively (see, e.g., Eq. (66) in Ref. \cite{BarJPhysG} and Eq. (67) in Ref. \cite{PRC2007}). To provide a resonance, the perturbing spin-tensor interaction should oscillate with the angular frequency $\omega\approx2\Omega'$ \cite{Bar3,BarJPhysG,PRC2007}. The same situation takes place for the intrinsic OAM. It can be similarly shown that the resonance frequency of the perturbing tensor interaction (10) is $\omega=2\Omega$. The perturbing nonuniform quasielectric field is given by $\mathcal{E}_r(R,t)=\mathcal{E}_r(R)\cos{(\omega t+\varphi)}$. The quantity $\varphi$ is an oscillation phase at the moment of time $t=0$ when the initial beam polarization is defined by Eqs. (\ref{eq3}), (\ref{eq4}), and (\ref{eqPz}).
In the general case of an imperfect resonance, the dynamics of $P_z$ is defined by Eqs. (51) and (57) in Ref. \cite{PRC2007} and is given by
\begin{equation}
\begin{array}{c}
P_{z}(t)=\left(1-\frac{2{\cal
A}^2}{{\omega'}^2}\sin^2{\frac{\omega't}{2}}\right)\cos{\theta}+
\frac{{\cal
A}}{\omega'}\sin^2{\theta}\sin{\frac{\omega't}{2}}\biggl[\frac{2\Omega-\omega}{\omega'}\sin{\frac{\omega't}{2}}\cos{(2\psi-\varphi)}\\+\cos{\frac{\omega't}{2}}
\sin{(2\psi-\varphi)} 
\biggr]
\end{array}
\label{provg}
\end{equation}
and \cite{footnote}
\begin{equation}
\begin{array}{c}
P_{z}(t)=\frac{{\cal
A}}{\omega'}\sin^2{\theta}\sin{\frac{\omega't}{2}}\Bigl[\frac{2\Omega-\omega}{\omega'}\sin{\frac{\omega't}{2}}\cos{(2\psi-\varphi)}+\cos{\frac{\omega't}{2}}
\sin{(2\psi-\varphi)} 
\Bigr]
\end{array}
\label{propg}
\end{equation}
when the initial beam is vector-polarized and tensor-polarized, respectively. Here 
$$
\omega'=\sqrt{(2\Omega-\omega)^2+{\cal A}^2}. 
$$ 
The resonance part of the tensor interaction is defined by
\begin{eqnarray}
{\cal E}_r(R,t)&=&\mathfrak{G}(R-R_0)\cos{(\omega t+\varphi)},\qquad \nonumber \\
W&=&-\dfrac{Q_s\mathfrak{G}}{4L^2}\cos{(\omega t+\varphi)}L_r^2=2{\cal A}\cos{(\omega t+\varphi)}L_r^2,\nonumber \\
{\cal A}&=&-\dfrac{Q_s\mathfrak{G}}{8L^2}.
\label{defnA}
\end{eqnarray}
The definitions of ${\cal A}$ in Eqs. (\ref{defqA}) and (\ref{defnA}) are different.

The third method of the measurement of the EQM is very similar to the nuclear quadrupole resonance method. However, the nuclear quadrupole resonance takes place in the rest frame of the moving electron.

\end{document}